\documentclass[fleqn,usenatbib]{mnras}

\usepackage{newtxtext,newtxmath}

\usepackage[T1]{fontenc}

\DeclareRobustCommand{\VAN}[3]{#2}
\let\VANthebibliography\thebibliography
\def\thebibliography{\DeclareRobustCommand{\VAN}[3]{##3}\VANthebibliography}

\usepackage{graphicx}	
\usepackage{amsmath}	
\usepackage{siunitx}    
\usepackage{hyperref}   
\usepackage{multirow, makecell}   
\usepackage[table]{xcolor}     
\usepackage{color, soul}

\title[Dynamical Dark Energy in DESI DR2]{Comparison of dynamical dark energy with $\Lambda$CDM in light of DESI DR2}

\author[A.N.~Ormondroyd et al.]{
    A.N.~Ormondroyd,$^{1,2}$\thanks{E-mail: ano23@cam.ac.uk}
    W.J.~Handley,$^{2,3}$
    M.P.~Hobson$^{1}$
    and A.N.~Lasenby$^{1,2}$
    \\
    $^{1}$Astrophysics Group, Cavendish Laboratory, J.J.~Thomson Avenue, Cambridge, CB3 0HE, UK\\
    $^{2}$Kavli Institute for Cosmology, Madingley Road, Cambridge, CB3 0HA, UK\\
    $^{3}$Institute of Astronomy, Madingley Road, Cambridge, CB3 0HA, UK\\
    }

    \date{Accepted XXX. Received YYY; in original form ZZZ}

    \pubyear{2025}

\begin{document}
    \label{firstpage}
    \pagerange{\pageref{firstpage}--\pageref{lastpage}}
    \maketitle

    \begin{abstract}
        We present an updated reconstruction of the dark energy equation of state, $w(a)$, using the newly released DESI DR2 Baryon Acoustic Oscillation (BAO) data in combination with Pantheon+ and DES5Y Type Ia supernovae measurements, respectively.
        Building on our previous analysis in \cite{2025arXiv250308658O}, which employed a nonparametric flexknot reconstruction approach, we examine whether the evidence for dynamical dark energy persists with the improved precision of the DESI DR2 dataset. We find that while the overall qualitative structure of $w(a)$ remains consistent with our earlier findings, the statistical support for dynamical dark energy is reduced when considering DESI DR2 data alone, particularly for more complex flexknot models with higher numbers of knots.
        However, the evidence for simpler dynamical models, such as $w$CDM and CPL (which correspond to $n=1$ and $n=2$ knots respectively), increases relative to $\Lambda$CDM with DESI DR2 alone, with CPL being the preferred dynamical model, consistent with previous DESI analyses.
        When combined with Pantheon+ data, the conclusions remain broadly consistent with our earlier work, but when instead combined with DES5Y supernovae data, there is an increased preference for flexknot models for all values of $n$ considered. This results in all such models being preferred over $\Lambda$CDM, with the CPL model being the most favoured by a Bayes factor of $\sim 2.3$ relative to $\Lambda$CDM.
    \end{abstract}




    \section{Introduction}

    The standard model of cosmology, $\Lambda$CDM, has been remarkably successful in explaining a wide range of cosmological observations.

    Recent work has reinforced the importance of understanding the nature of dark energy through increasingly precise cosmological measurements.
    In our previous study \citep{2025arXiv250308658O}, we employed a nonparametric flexknot reconstruction \cite[originally termed `nodal reconstruction'][]{pkvazquez,devazquez} of the dark energy equation-of-state parameter, $w(a)$, to explore the possibility of dynamical dark energy.
    Using a flexible linear-spline approach with free-moving nodes, our analysis of DESI Baryon Acoustic Oscillation (BAO) combined with either Pantheon+ or DES5Y Type Ia supernovae data unexpectedly revealed a W-shaped structure in $w(a)$.
    This structure, which deviates from the conventional constant-$w$ ($\Lambda$CDM) picture, raised questions about whether standard parameterisations such as $w$CDM or CPL might be too restrictive to capture the true dynamical behavior of dark energy.
    This is acknowledged in the DESI DR2 release \citep{desi2025, desi2i, desi2ii}, which includes an entire paper dedicated to an extended dark energy analysis \citep{desi2de}.

    In this update, we investigate how the conclusions of \cite{2025arXiv250308658O} change in light of DESI DR2.

    \vfill

    \section{Data}\label{sec:data}

    \begin{figure}
        \centering
        \includegraphics[width=0.48\textwidth]{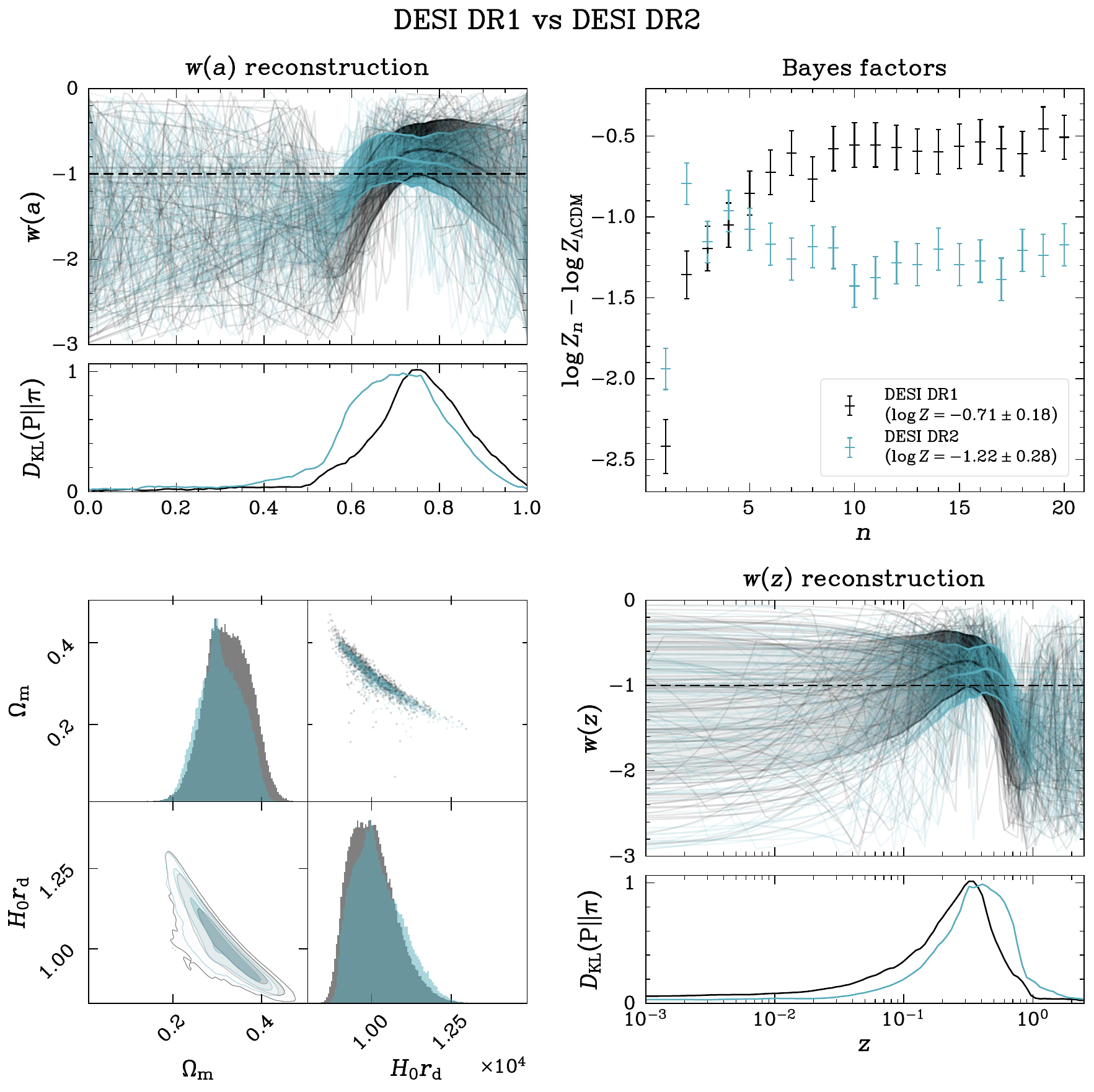}
        \caption{
            Flexknot reconstruction of $w(a)$ using DESI DR2 BAO data using flexknots, compared to DR1.
            Upper left panel: the reconstructed $w(a)$.
            The dashed line is the mean of the posterior, and the shaded region is the $1\sigma$ contour. The opacity of the shaded region is proportional to the KL divergence of $w(a)$.
            The overall shape is similar to DR1, but the transition around $a=0.6-0.7$ is less pronounced.
            Below, the KL divergence of $w$ is shown as a function of $a$.
            Upper right panel: the evidence for each number of knots $n$, relative to $\Lambda$CDM.
            Evidence for models with one or two knots is increased compared to DR1, whereas evidence for models with more than four knots is reduced.
            Lower left panel: posterior distributions of $\Omega_\mathrm m$ and $H_0r_\mathrm d$.
            Lower right panel: the same reconstruction as the upper left panel, but transformed to $w(z)$.
            The KL divergence is similarly shown as a function of $z$.
        }\label{fig:desidr12}
    \end{figure}

    \begin{figure}
        \begin{center}
            \includegraphics[width=0.48\textwidth]{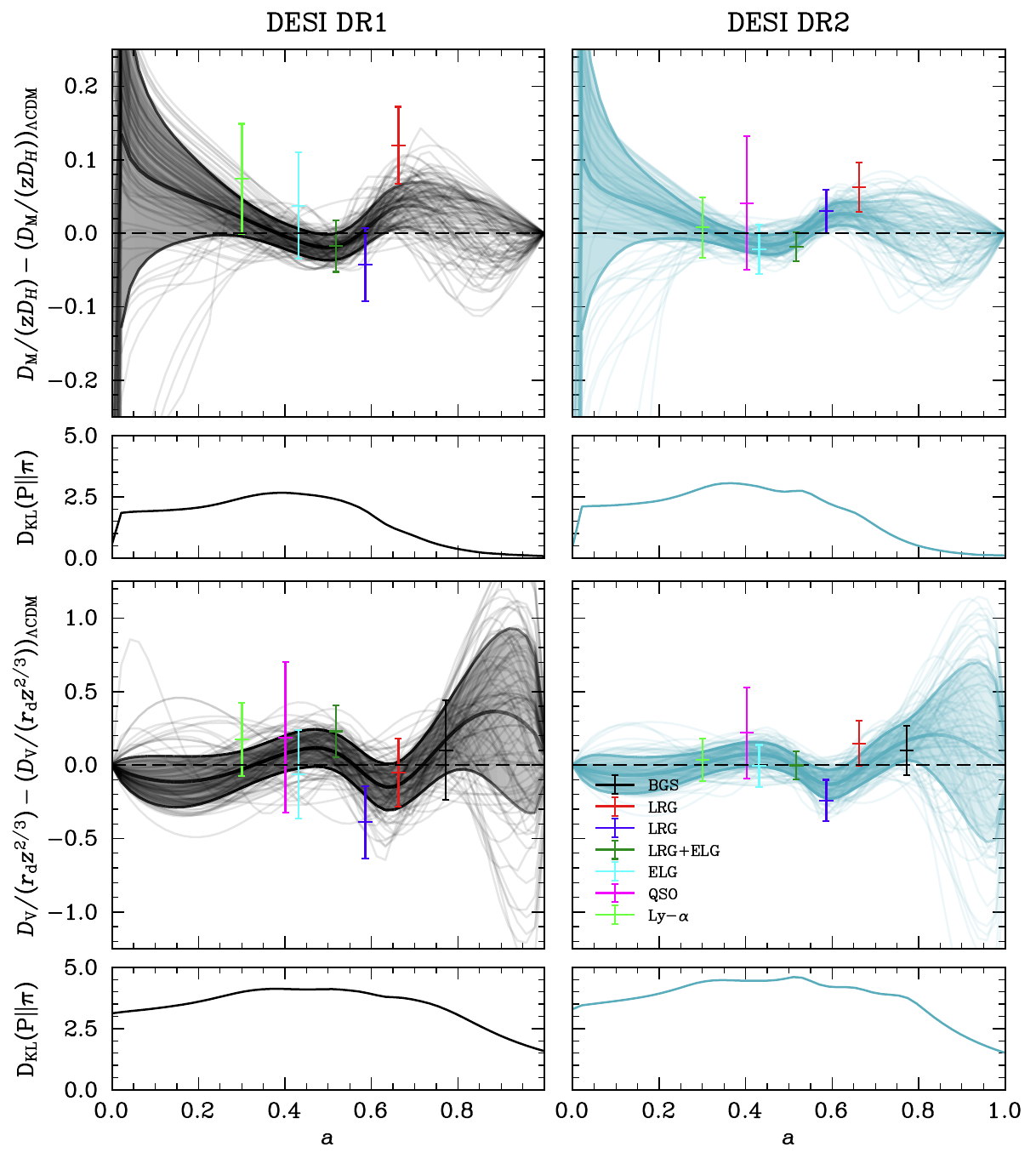}
        \end{center}
        \caption{
            BAO distances reconstructed from DESI BAO data.
            The left column is DESI DR1, the right column is DESI DR2.
            The best-fit $\Lambda$CDM has been subtracted from each set of distances.
            The KL divergence of the distances is shown as a function of $a$ below each reconstruction.
            $1\sigma$ contours are also shown, and their opacity is proportional to the KL divergence.
            It can been seen how the smaller error bars in DR2 have produced a narrower $1\sigma$ contour for the reconstructions, which corresponds to the greater peak KL divergence.
        }\label{fig:distances}
    \end{figure}

    We combine DESI DR2 BAO data with Pantheon+ \citep{pantheonplus} and DES5Y supernovae \citep{des5y}, respectively.
    DESI DR2 cosmological distances are used as they appear in Table~IV of \cite{desi2ii}.
    DESI DR1, Pantheon+, and DES5Y data 
    are used in precisely the same manner as in \cite{2025arXiv250308658O}.

    The DESI DR2 BAO improves constraints on cosmic expansion with a larger dataset of galaxies and quasars than DR1.
    In addition to the tightening of the error bars compared to the previous release, the Quasar Sample (QSO) now has sufficient signal-to-noise ratio that separate measurements of $D_\mathrm M(z)$ and $D_\mathrm H(z)$ are reported in DR2, whereas in DR1 only a volume-averaged $D_\mathrm V(z)$ value was reported \citep{desi2ii, desiiii}.

    \section{Methods}

    With these improved DESI DR2 measurements now available, we recap the flexknot-based methodology that enables us to explore dynamical dark energy in a nonparametric way.
    In this approach, $w(a)$ is modelled using a flexible linear spline between free-moving nodes.
    This technique is well-established in multiple fields within cosmology: it has been used to reconstruct history of the dark energy equation of state from CMB data \citep{sonke, devazquez}, the primordial power spectrum \citep{pkhandley, pkvazquez, pkknottedsky, pkcore, pkplanck13, pkplanck15}, the cosmic reionisation history \citep{flexknotreionization, heimersheimfrb}, galaxy cluster profiles \citep{flexknotclusters}, and the $\SI{21}{\centi\metre}$ signal \citep{heimersheim21cm, shen}.
    Unlike dark energy reconstructions such as Gaussian processes \citep{modelagnosticgp, quintom, 2025arXiv250304273J, 2025arXiv250315943G} or cubic splines \citep{2025arXiv250313198B}, flexknots can reconstruct arbitrarily sharp features, and have extremely weak functional correlation structure. Flexknots also have the advantage that the $w$CDM and CPL models correspond to the special cases of $n=1$ and $n=2$ knots, respectively. 

    An addition to the figures compared to \cite{2025arXiv250308658O} is the KL divergence \citep{dkl} of the functional posterior, which is shown as a function of $a$ or $z$ below the reconstructed $w$.
    The opacity of the mean and $1\sigma$ contours are proportional to the KL divergence.

    \begin{table}
        \centering
        \rowcolors{2}{}{gray!25}
        \begin{tabular}{|l|c|}
            \hline
            Parameter & Prior \\
            \hline
            $n$ & $[1, 20]$ \\
            $a_{n-1}$ & $0$ \\
            $a_{n-2}, \dots, a_1$ & sorted($[a_{n-1}, a_0])$ \\
            $a_0$ & $1$ \\
            $w_{n-1}, \dots, w_0$ & $[-3, -0.01]$ \\
            $\Omega_\mathrm m$ & $[0.01, 0.99]$ \\
            $H_0r_\mathrm d$ (DESI)& $[3650, 18250]$ \\
            $H_0$ (Ia) & $[20, 100]$ \\
            \hline
        \end{tabular}
        \caption{
            Cosmological priors used in this work.
            Fixed values are indicated by a single number, while uniform priors are denoted by brackets.
            As BAO only depend on the product $H_0r_\mathrm d$, and supernovae depend on $H_0$, those parameters are only included as necessary. Whilst the dynamical dark energy priors are broadly consistent with those in \protect\cite{desi2ii}, these inevitably differ from CPL priors which instead put a uniform prior on the gradient $w_a$.
        }
        \label{tab:priors}
    \end{table}

    \begin{figure*}
        \centering
        \includegraphics[width=0.48\textwidth]{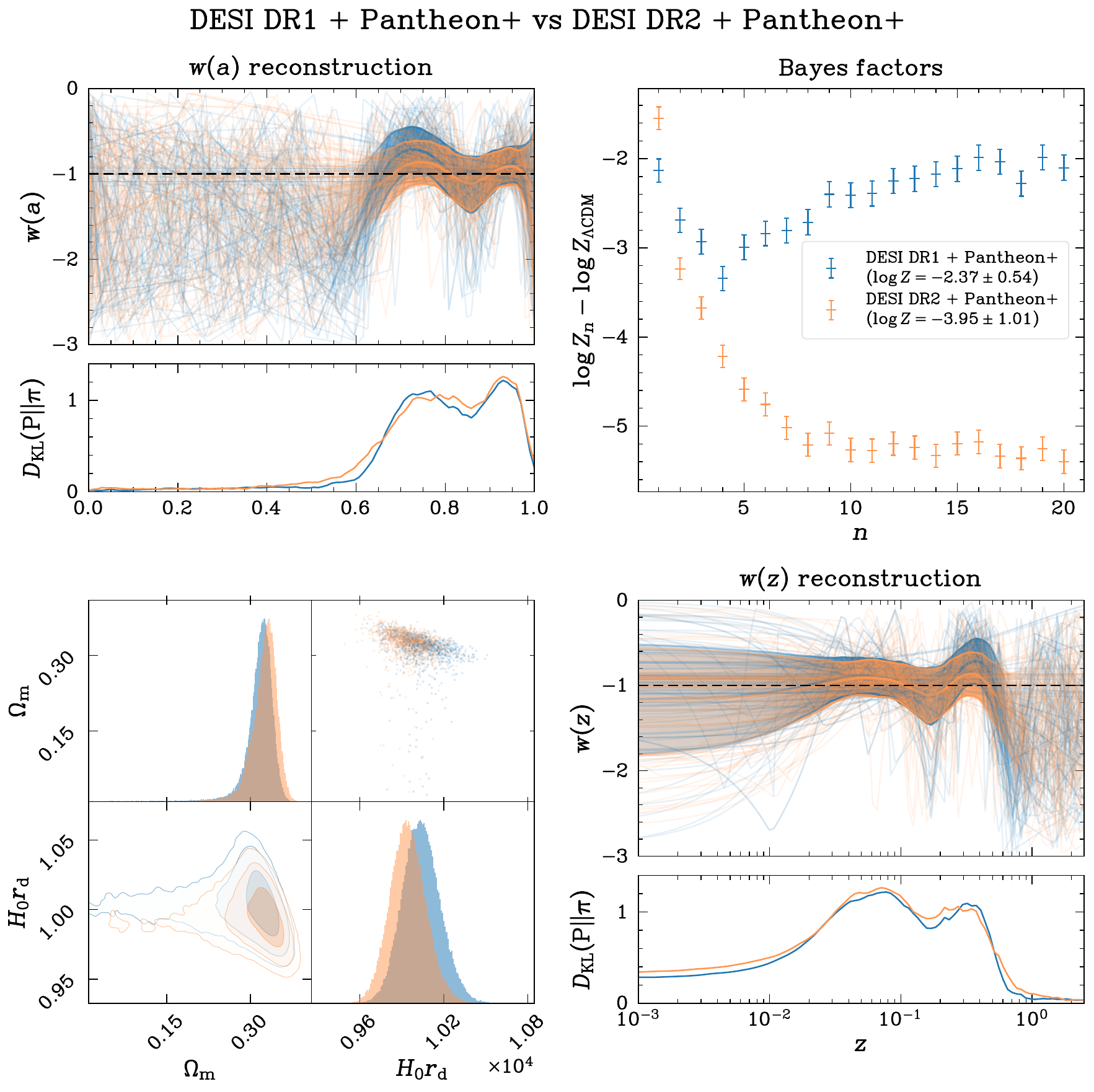}
        \includegraphics[width=0.48\textwidth]{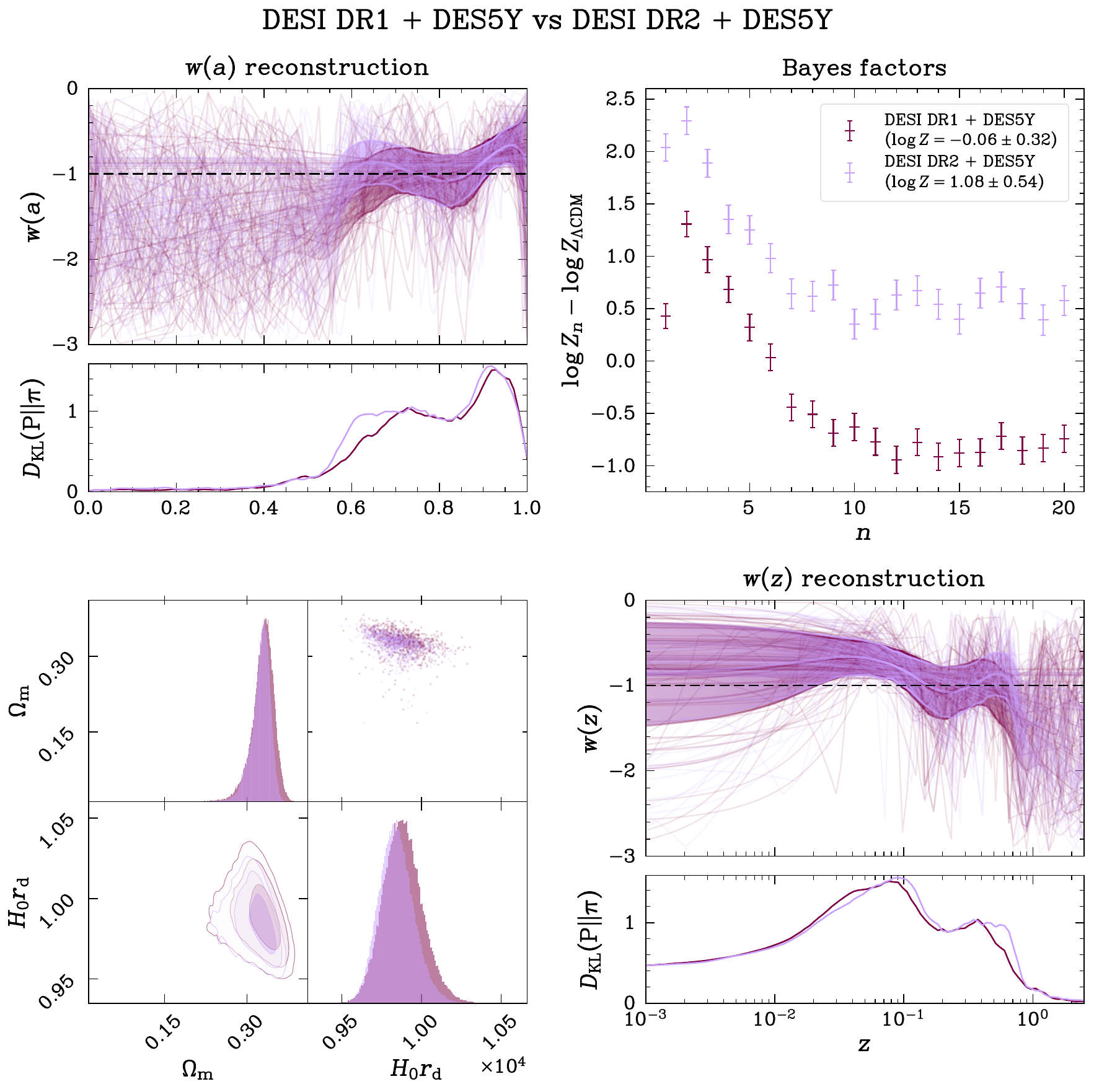}
        \caption{
            Similar to Figure~\ref{fig:desidr12}, but comparing DESI DR2 BAO data with Pantheon+ supernovae (left) and DES5Y supernovae (right).
            The most significant difference between the DR1 and DR2 reconstructions is the same region of $a$ as in Figure~\ref{fig:desidr12}.
            Left four panels: DESI DR2 + Pantheon+, compared to DR1 + Pantheon+.
            The evidence for $w$CDM is very similar between DR1 and DR2, but the remaining flexknots are less favoured in DR2.
            Right four panels: DESI DR2 + DES5Y, compared to DR1 + DES5Y.
            The change is very much the opposite as it was with Pantheon+, with the evidence for all numbers of knots greater than or equal to two have increased significantly, with CPL remaining the preferred model.
        }\label{fig:desidr2ia}
    \end{figure*}

    \begin{figure}
        \begin{center}
            \includegraphics[width=0.48\textwidth]{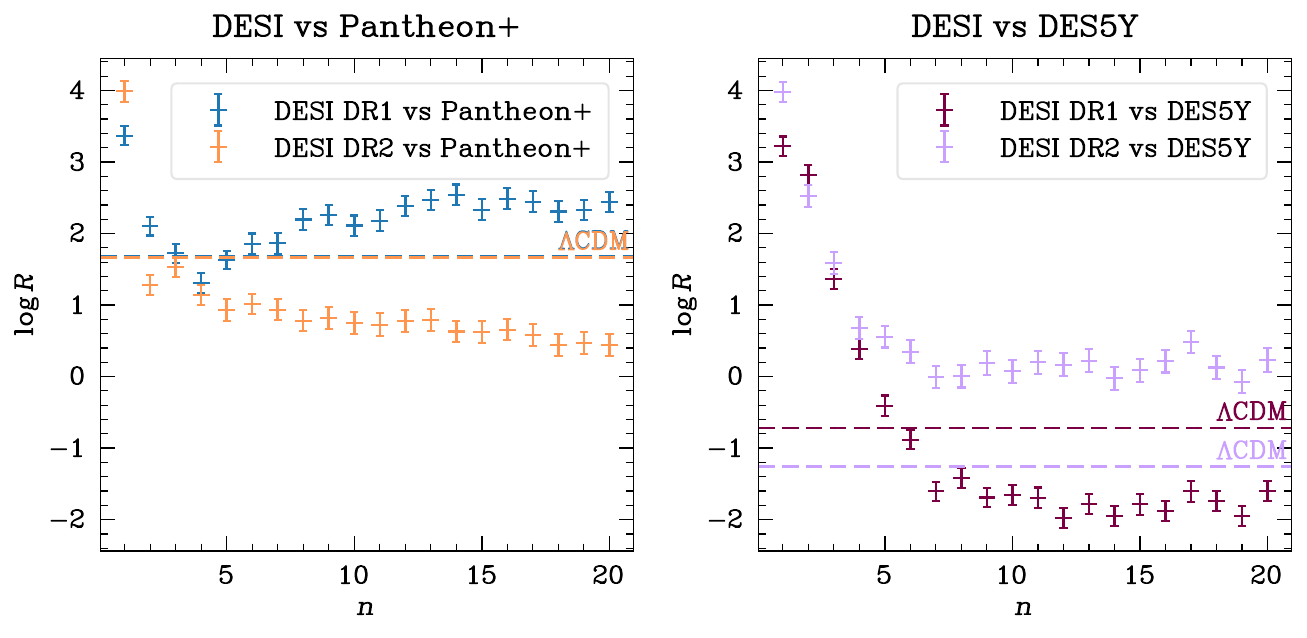}
        \end{center}
        \caption{
            Tension quantifications between combinations of datasets.
            For each knot reconstruction $n$ we compare the tension between DESI (DR1 or DR2) and supernovae (Pantheon+ [left panel] or DES5Y [right panel]).
            The $\log R$ value for $\Lambda$CDM for each dataset pair are shown by the dashed horizontal lines.
            Note that for DESI vs Pantheon+, this value is almost unchanged between DR1 and DR2 so the lines are on top of each other.
            In light of the update from DR1 to DR2, the tension has increased between DESI and Pantheon+ ($\log R$ lower), but remains consistent ($\log R > 0$).
            The exception is $n=1$ ($w$CDM), which has reduced in tension, but is still the model with the least tension.
            For DESI and DES5Y in light of the update from DR1 to DR2 the tension has decreased ($\log R$ higher) and is now consistent ($\log R>0$) for all but the $\Lambda$CDM ($N=0$) case.
        }\label{fig:tension}
    \end{figure}

    Posterior samples and evidences were obtained using the nested sampling algorithm \texttt{PolyChord} \citep{skilling2004, polychord1, polychord2}.
    A branch of \texttt{fgivenx} was used to produce the functional posterior plots \citep{fgivenx}, and \texttt{anesthetic} was used to process the nested sampling chains \citep{anesthetic}.

    As in \cite{2025arXiv250308658O}, $H_0$ and $M_\mathrm B$ are marginalised over, and the prior on $M_\mathrm B$ is taken to be uniform and sufficiently wide to contain the entire posterior.
    In the DESI analyses \citep{desi2ii, desivi}, it is enforced that $w_0+w_a<0$ to ensure that there is a period of matter domination at high redshifts.
    In this work, we take $w_i < 0$ to achieve the same effect.

    Table~\ref{tab:priors} lists the cosmological priors used in this work, the same as \cite{2025arXiv250308658O}, which themselves were chosen to remain consistent with those used in \cite{desicrossingstatistics}.
    \pagebreak

    \section{Results}\label{sec:results}
    
    Figure~\ref{fig:desidr12} shows  reconstructions of $w(a)$ from DESI alone, and compares DR1 and DR2.
    Figure~\ref{fig:distances} shows the corresponding BAO distances reconstructed from DESI BAO data, compared to the best-fit $\Lambda$CDM model.
    The most significant change is not in the shape of the $w(a)$ posterior, but the evidences for each number of knots $n$ in the reconstruction.
    As the number of knots increases, the evidences tend to a lower value for DR2 than for DR1.
    This means that DR2 is less supportive than DR1 of the flexknot model.

    The most significant change to the reconstructed $w(a)$ is the transition around $a=0.6-0.7$, which is the region containing the two LRG points whose effect was previously investigated in detail \citep{2025arXiv250308658O}.
    This has moved to slightly higher redshifts, and is tighter than in DR1.
    The DR2 reconstructions have a more pronounced transition at slightly higher redshifts between phantom and quintessence than DR1, which continues to be the case when supernovae are included.
    This is echoed in the corresponding functional KL divergence for $w(a)$, which remains around $1$ for greater redshifts in DR2 than DR1, though both tend to zero below $a\approx 0.5$.

    Looking specifically, however, at $n=1$ and $n=2$ (i.e. the $w$CDM and CPL parameterisations), we note that the evidences slightly increase relative to $\Lambda$CDM with DR2 compared to DR1.
    This matches the conclusion of \cite{desi2ii} that DR2 more strongly supports these dark energy models over $\Lambda$CDM than DR1, with CPL being the preferred dynamical model, although still having a marginally lower evidence than $\Lambda$CDM.
    In Figure~\ref{fig:distances} we see that the BAO distances reconstructed from DESI DR2 are broadly similar to those from DR1, but with narrower $1\sigma$ contours, which correspond to the marginally greater peak KL divergences.

    Figure~\ref{fig:desidr2ia} shows a comparison  of results from the DESI DR1/DR2 BAO data when combined with Pantheon+ and DES5Y supernovae, respectively.
    The evidence for $w$CDM in DR2 are greater than those with DR1 and, as when using DESI alone, the evidences tend to a lower value for DR2 than for DR1 as the number of knots $n$ increases.
    Most notably, there is now a preference for models with large numbers of knots with DESI DR2 + DES5Y, which was the only combination in \cite{2025arXiv250308658O} which had any evidence in favour of dynamical dark energy, and the CPL model is still favoured.
    
    When combining data it is prudent to check that the datasets are consistent with each other. Here we follow \cite{2025arXiv250308658O} in using the tension analysis developed and deployed in \cite{lemos, hergt, balancingact} via the $\log R$ statistic. 
    In Figure~\ref{fig:tension} we show the tension quantifications between DESI and supernovae datasets.
    For DESI vs Pantheon+, there has been a slight increase in tension between DR1 and DR2, but they remain consistent.
    The exception is $n=1$ ($w$CDM), which has reduced in tension, but was already the model with the least tension.
    For DESI vs DES5Y the tension has decreased between DR1 and DR2, and is now consistent for all but the $\Lambda$CDM case.
    In this sense, dynamical dark energy models can be viewed as resolving a discrepancy between DESI and DES5Y data, in addition to providing a better fit to the data.

\begin{figure*}
        \begin{center}
            \includegraphics[width=0.95\textwidth]{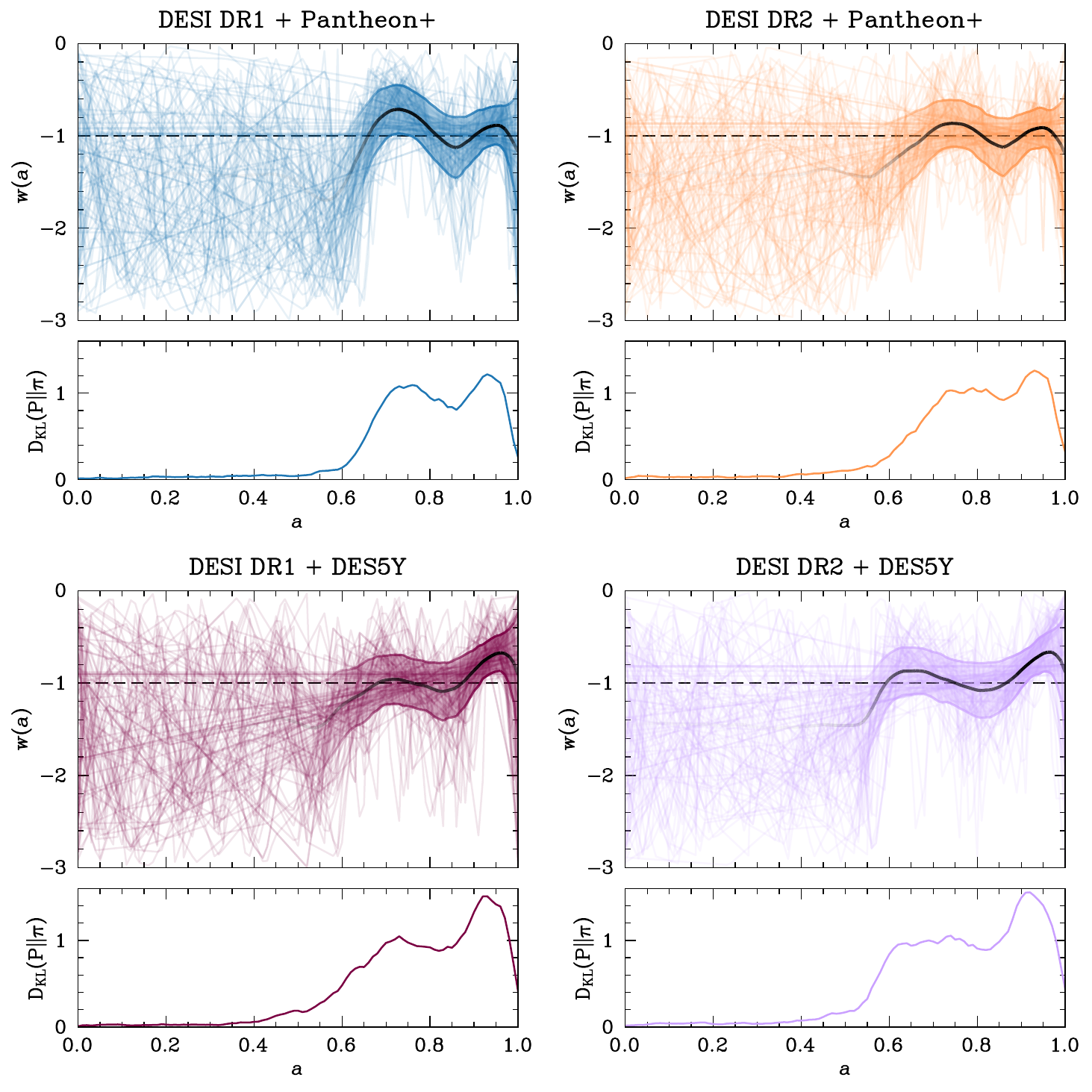}
        \end{center}
        \caption{
            For reference, non-overlaid reconstructions of $w(a)$ for the remaining combinations of datasets. The KL divergence of $w(a)$ is shown below each reconstruction.
        }\label{fig:separated}
    \end{figure*}

    For reference, we show the non-overlaid reconstructions of $w(a)$ for the remaining combinations of datasets in Figure~\ref{fig:separated}.

    \section{Conclusions}\label{sec:conclusions}

    We revisited the dynamical dark energy reconstructions presented in \cite{2025arXiv250308658O} using the newly-released DESI DR2 BAO data in combination with Pantheon+ and DES5Y supernovae measurements, respectively.
    Our analysis employed a flexknot methodology to reconstruct the evolution of the dark energy equation of state $w(a)$.
    Overall, we find that while the qualitative shape of the reconstructed $w(a)$ remains consistent with our previous work, there is a marked change in the statistical evidence for dynamical dark energy.

    The DESI DR2 data alone lead to a reduction in the evidence as compared with DR1 for flexknot models with larger numbers of knots.
    However, the evidence for the $w$CDM and CPL models, which correspond to $n=1$ and $n=2$ knots respectively, have increased compared to $\Lambda$CDM with DESI alone, which aligns with the conclusions of \cite{desi2ii}.
    When the DESI DR2 BAO data are combined with Pantheon+ supernovae, the conclusions are similar to those in our original work.
    
    However, with DES5Y supernovae, there is now increased evidence for models with a larger number of knots, with evidence for CPL, which remains the preferred model, also increasing, and with $w$CDM now also favoured over $\Lambda$CDM. In addition to providing a better fit, dynamical dark energy models serve to resolve a discrepancy between DESI and DES5Y data.

    \section*{Acknowledgements}

    This work was performed using the Cambridge Service for Data Driven Discovery (CSD3), part of which is operated by the University of Cambridge Research Computing on behalf of the STFC DiRAC HPC Facility (\url{www.dirac.ac.uk}).
    The DiRAC component of CSD3 was funded by BEIS capital funding via STFC capital grants ST/P002307/1 and ST/R002452/1 and STFC operations grant ST/R00689X/1.
    DiRAC is part of the National e-Infrastructure.
    WJH was supported by a Royal Society University Research Fellowship.

    The tension calculations in this work made use of \texttt{NumPy} \citep{numpy}, \texttt{SciPy} \citep{scipy}, and \texttt{pandas} \citep{pandaszenodo, pandaspaper}.
    The plots were produced in \texttt{matplotlib} \citep{matplotlib}, using the \texttt{smplotlib} template created by \citet{smplotlib}.

    \section*{Data Availability}

    The pared-down Python pipeline and nested sampling chains used in this work can be obtained from Zenodo \citep{ormondroyd_2025_15025604}.



    \bibliographystyle{mnras}
    \bibliography{desi2} 




    \appendix

    \section{Effect of flexknot prior}

    \begin{figure*}
        \begin{center}
            \includegraphics[width=0.95\textwidth]{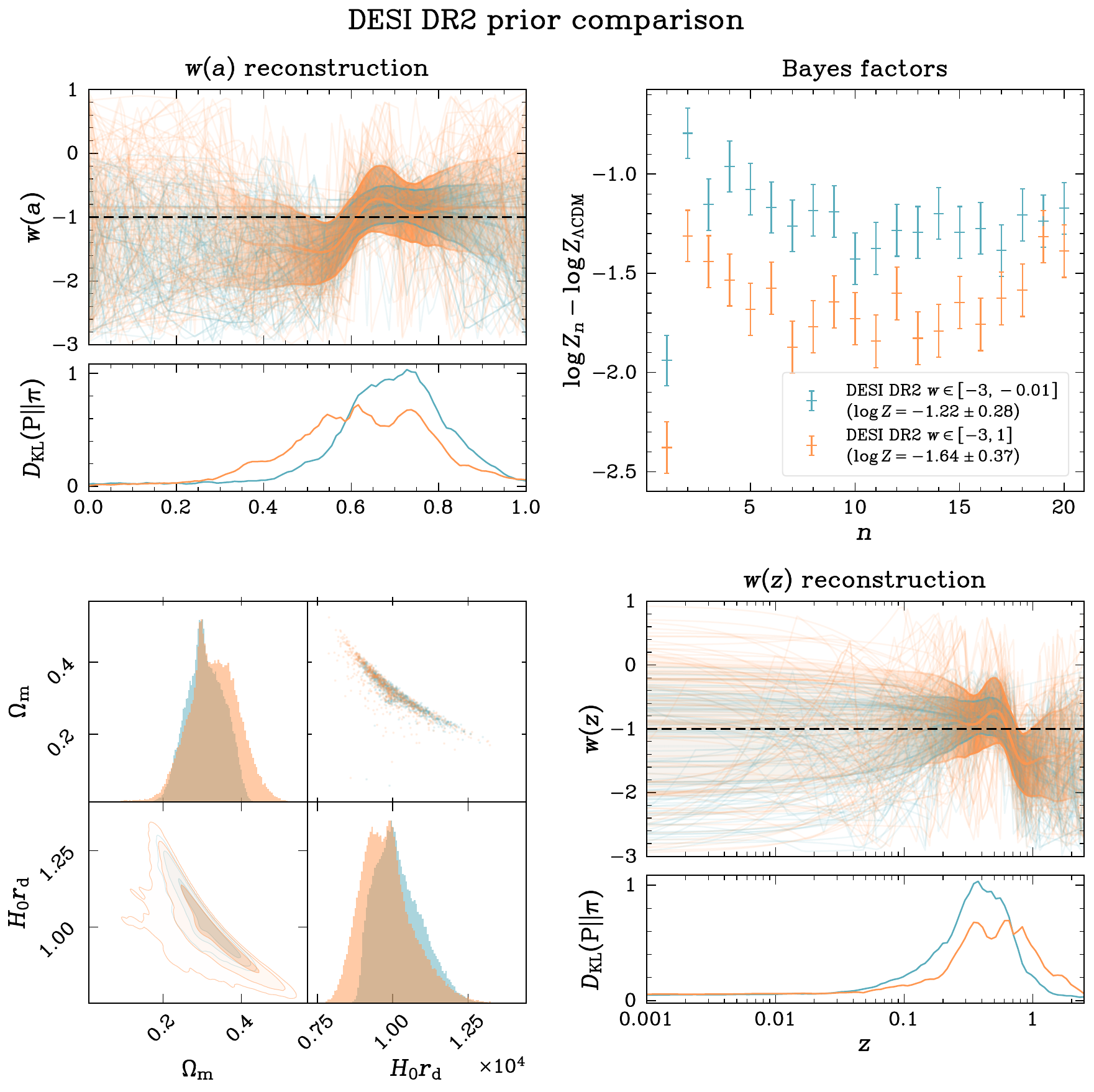}
        \end{center}
        \caption{
            Two flexknot reconstructions of $w(a)$ using DESI DR2 BAO data, using different priors on the $w$ parameters: $[-3, -0.01]$ (black) and $[-3, 1]$ (orange).
            The two reconstructions are very similar at low redshifts, but the transition from quintessence to phantom behaviour is more abrupt with the wider prior.
            Below $a\approx0.4$, the posteriors return to their respective priors, as expected.
            Interestingly, the return is quite gradual for the mean of the reconstruction, which is more noticeable with the wider prior as the structure finishes at approximately $w~-1.5$, i.e. the centre of the narrower prior.
            The trends in the evidences for each number of knots $n$ are similar, so the conclusions of this work are unaffected by this alternative prior choice.
        }\label{fig:wide}
    \end{figure*}

    The Bayesian framework used in this work is, of course, dependent on the choice of prior.
    CPL parameterisations ($w(a) = w_0 + (1-a)w_a$) of dark energy typically use a uniform prior on $w_0$ (the value of $w$ at $a=1$) of $[-3, 1]$, \citep[e.g.][]{desiiii, desi2ii, desi2de}, which is the value of $w$ in $\Lambda$CDM.
    This is perhaps motivated by the fact that the centre of the uniform prior is $-1$, which is the value of $w$ in $\Lambda$CDM.
    However, this notion is flawed on two counts: firstly: a uniform prior is precisely that, uniform, and so the centre of the prior is not special, in the same way that there is no problem that the DESI prior for $w_a$ of $[-3, 2]$ is not centred on the $\Lambda$CDM value of 0. Secondly, there is usually an additional constraint that $w_0 + w_a < 0$ to ensure that there is a period of matter domination at high redshifts, which cuts off the upper-right corner of the joint prior on $w_0$ and $w_a$.

    In this work, we have used a prior of $[-3, -0.01]$ for the $w$-coordinates of the flexknot parameters (see Table~\ref{tab:priors}), which is the same as in \cite{2025arXiv250308658O}, therefore ``centred'' on $-1.505$.
    This achieves the same guarantee of a period of matter domination at high redshifts.
    Unconstrained regions of $w$ of the flexknot reconstructions will return the prior, i.e. a mean of approximately $-1.5$ with a $1\sigma$ uncertainty of $3/2\sqrt{3} \approx 0.86$, which is consistent the left-hand regions of the reconstructions throughout the main paper.
    Clearly, they should not be interpreted as a preference for $w \approx -1.5$, perhaps this is a flaw of plotting the mean and $1\sigma$ contours which are intended to guide the eye through the constrained regions of the functional posterior.

    Nevertheless, we have checked that the conclusions of this work are unchanged if we instead use a prior of $[-3, 1]$ for the $w$ parameters in the flexknot reconstruction.
    For example, Figure~\ref{fig:wide} shows the flexknot reconstruction of $w(a)$ using a prior of $[-3, 1]$ with DESI DR2 data, overlaid atop the reconstruction with the prior of $[-3, -0.01]$, which was shown previously in the left panel of Figure~\ref{fig:desidr2ia} and the upper-right panel of Figure~\ref{fig:separated}.

    The two posteriors are very similar at low redshifts, though the highest-redshift transition from quintessance to phantom behaviour is more abrupt.
    Below $a\approx0.4$, the posteriors return to their respective priors, as expected.
    Interestingly, the return is quite gradual for the mean, which is more noticeable with the wider prior as the structure finishes at approximately $w~-1.5$, i.e. the centre of the narrower prior.
    This is a feature of the flexknot approach, which penalises structure where it is not required by the data, and a gradual change requires fewer knots.
    Crucially, the evidences for each number of knots show similar trends, so the conclusions of this work are unchanged by this reasonable alternative choice of prior.

    \section{Simplified $H_0$ marginalisation}

    Since the publication of \cite{2025arXiv250308658O}, we have realised that the result can be simplified.
    After marginalising over $M_\mathrm B$ in Appendix~B1 of that work, we were left with:
    \begin{equation}
        \begin{aligned}
            \mathcal L &= \frac{1}{V_{M_\mathrm B}} \sqrt{\frac{2\pi}{|2\pi\Sigma|\mathbfit 1^T\Sigma^{-1}\mathbfit 1}} \exp - \frac 1 2 \mathbfit x^T\tilde\Sigma^{-1}\mathbfit x \text,\\
            \tilde\Sigma^{-1} &= \Sigma^{-1} - \frac{\Sigma^{-1}\mathbfit 1\mathbfit 1^T\Sigma^{-1}}{\mathbfit 1^T\Sigma^{-1}\mathbfit 1} \text, \quad \mathbfit x = \mathbfit m_\mathrm B - \mu(\mathbfit z, \theta) \text.
        \end{aligned}
    \end{equation}
    Like before, the dependence on $H_0$ of $\mathbfit x$ is separated out:
    \begin{equation}
        \begin{aligned}
            x &= \mathbfit m_\mathrm B - \mu(\mathbfit z, \theta) = \mathbfit m_\mathrm B - 5\log_{10}{\frac{D_\mathrm L(\mathbfit z)}{\SI{10}{pc}}}\\
            &= \mathbfit m_\mathrm B - 5\log_{10}{(1+\mathbfit z_\mathrm{hel})\int_0^{\mathbfit z_\mathrm{HD}}\frac{\mathrm dz'}{h(z')}} - 5\log_{10}{\left(\frac{c}{\SI{10}{pc}H_0}\right)}\\
            &= 5\log_{10}\left(\frac{\SI{10}{pc}H_0}{c}\right) - \mathbfit y = h - \mathbfit y\text.
        \end{aligned}
    \end{equation}

    Let $h$ be itself multiplied by a vector of ones, and plug this back into the argument of the exponential:
    \begin{equation}
        - \frac 1 2 \mathbfit x^T\tilde\Sigma^{-1}\mathbfit x = - \frac 1 2 \mathbfit y^T\tilde\Sigma^{-1}\mathbfit y + h \mathbfit 1^T\tilde\Sigma^{-1}\mathbfit y - \frac 1 2 \mathbfit 1^T\tilde\Sigma^{-1}\mathbfit 1 h^2\text.
    \end{equation}

    Previously, we noted that $\mathbfit 1^T\tilde\Sigma^{-1}\mathbfit 1 = 0$, however, we should also have noted that $\mathbfit 1^T\tilde\Sigma^{-1} = \mathbfit 0^T$. Thus, the dependence of the likellihood on $H_0$ vanishes.
    This has no effect on the posterior, as the additional terms in the original version were zero within numerical error.
    However, it is interesting to note that this means that the evidence and the posterior on the other parameters is completely independent of the prior on $H_0$.


    \bsp	
    \label{lastpage}
\end{document}